\documentclass{article}
\usepackage{arxiv}
\usepackage[utf8]{inputenc} 
\usepackage[T1]{fontenc}    
\usepackage{hyperref}       
\usepackage{url}         
\usepackage{booktabs}    
\usepackage{amsfonts}      
\usepackage{nicefrac}      
\usepackage{microtype}     
\usepackage{lipsum}
\usepackage{float}
\usepackage{graphicx}
\usepackage{caption}
\usepackage{subcaption}
\usepackage{titlesec}
\usepackage{multirow}
\usepackage{amsmath}
\usepackage{soul,color}
\usepackage{multicol}
\usepackage{placeins} 

\usepackage[nodisplayskipstretch]{setspace}

\title{ MLatticeABC: Generic Lattice Constant Prediction of Crystal Materials using Machine Learning}

\author{
 Yuxin Li, Wenhui Yang, Rongzhi Dong\\
 School of Mechanical Engineering\\
  Guizhou University \\
  Guiyang China 550025 \\
    \And
    Jianjun Hu*\\
  Department of Computer Science and Engineering\\
  University of South Carolina\\
  Columbia, SC 29201 \\
\texttt{jianjunh@csc.sc.edu} \\
}

\titlespacing {\section}
{0pt}{5.5ex plus 1ex minus .2ex}{3.3ex plus .2ex}
\begin{document}
\maketitle
\begin{abstract}
Lattice constants such as unit cell edge lengths and plane angles are important parameters of the periodic structures of crystal materials. Predicting crystal lattice constants has wide applications in crystal structure prediction and materials property prediction. Previous work has used machine learning models such as neural networks and support vector machines combined with composition features for lattice constant prediction and has achieved a maximum performance for cubic structures with an average $R^2$ of 0.82. Other models tailored for special materials family of a fixed form such as ABX\textsubscript{3} perovskites can achieve much higher performance due to the homogeneity of the structures. However, these models trained with small datasets are usually not applicable for generic lattice parameter prediction of materials with diverse composition. Herein, we report MLatticeABC, a random forest machine learning model with a new descriptor set for lattice unit cell edge length ($a,b,c$) prediction which achieves a $R^2$ score of 0.979 for lattice parameter $a$ of cubic crystals and significant performance improvement for other crystal systems as well. Source code and trained models can be freely accessed at \url{https://github.com/usccolumbia/MLatticeABC}

\end{abstract}

\section{Introduction}



\begin{figure}[h]
	\centering
	\begin{subfigure}{.45\textwidth}
		\includegraphics[width=\textwidth]{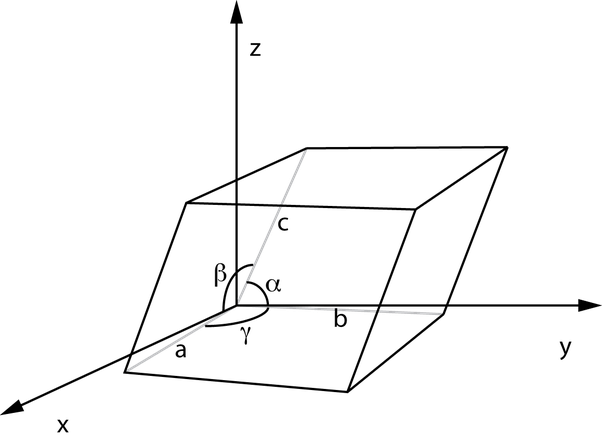}
		\caption{Unit cell and lattice constants: $a,b,c$ and $\alpha, \beta, \gamma$.}
		\vspace{3pt}
	\end{subfigure}
	\begin{subfigure}{.35\textwidth}
		\includegraphics[width=\textwidth]{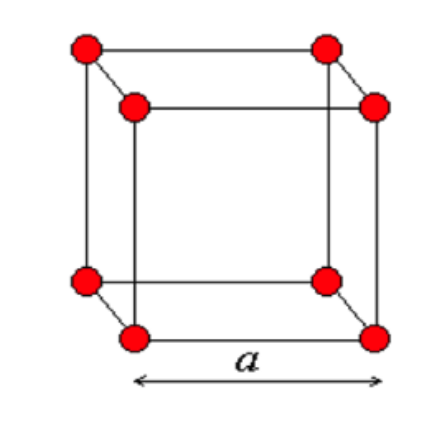}
		\caption{Cubic unit cell where $a=b=c$ and $\alpha=\beta=\gamma=90^o$}
		\vspace{3pt}
	\end{subfigure}

	\caption{Unit cell descriptors and lattice constants.}
	\label{fig:unitcell}
\end{figure}

 The periodic structures of crystal materials can be summarized by its space group and the parallelepiped unit cell as shown in Figure~\ref{fig:unitcell}. A unit cell is defined by six lattice parameters/constants including the lengths of three cell edges ($a, b, c$) and the angles between them ($\alpha, \beta, \gamma$). The shape of the unit cell of a crystal material determines its crystal system out of seven possibilities: triclinic, monoclinic, orthorhombic, tetragonal, trigonal, hexagonal, and cubic. Lattice constants and their changes such as lattice distortion upon different pressures and temperatures are related to many interesting physical and chemical properties of the materials\cite{mccaffrey1973electronic,pozzo2011thermal,wang2017lattice}. Lattice mismatch between the films and the growth substrates is also known to cause major issues in fabricating large and high quality of heteroepitaxial films of semiconductors like GaAs, GaN and InP \cite{jiang2006prediction}. Finding a new material with desired matched lattice constant is then a big challenge for both experimental approach based on X-ray electron or neutron diffraction techniques and first-principles calculations for large-scale screening.


Computational prediction of lattice constants of crystal materials has wide applications in both materials property prediction and discovery\cite{liang2020cryspnet}, crystal structure prediction\cite{hu2020contact,hu2020distance}, and large screening of materials for materials fabrication\cite{nait2020prediction}. During the past 15 years, a series of prediction approaches have been proposed for lattice constant prediction, which can be categorized by their input information used, the descriptors or features, the machine learning model, and the chemical system or materials family they are trained for. According to the input information of the prediction models, the approaches can be divided into composition (such as atomic properties of their constituent elements) based lattice parameter prediction models \cite{jiang2006prediction} and structure based prediction models\cite{williams2020deep}. While the majority of methods are based on composition information, the structure based approaches can also bring interesting insights \cite{nait2020prediction}. In \cite{williams2020deep}, a deep learning method is proposed to predict lattice parameters in cubic inorganic perovskites based on Hirshfeld surface representations of crystal structures. They showed that two-dimensional Hirshfeld surface fingerprints contain rich information encoding the relationships between chemical bonding and bond geometry characteristics of perovskites. 

Lattice prediction methods can also be classified by their machine learning models used. A variety of machine learning models have been applied for lattice constant prediction including conventional machine learning methods in the early stage such as linear regression \cite{chonghe2003prediction,jiang2006prediction}, support vector machines \cite{javed2007lattice,takahashi2017descriptors}, neural networks \cite{chonghe2003prediction,cao2019convolutional}, gene expression programming\cite{nait2020prediction}, random forests(RF)\cite{majid2011predicting}, and Gaussian process regression (GPR)\cite{zhang2020machine}. However, the performance difference among different machine learning algorithms have not been thoroughly evaluated.

A major difference among different lattice constant prediction studies is the chemical systems or materials families they focus on. The majority of the studies are focused on a special category of materials with fixed mole ratios, including the cubic perovskites ABX\textsubscript{3}\cite{jiang2006prediction,moreira2007comment,javed2007lattice,majid2010lattice,ubic2015lattice,ubic2009prediction}, double perovskites\cite{majid2009lattice}, A\textsubscript{2}XY\textsubscript{6} \cite{nait2020prediction},half-Heusler ternary compounds(XYZ)\cite{ahmad2019artificial}, and binary body centered cubic crystals \cite{takahashi2017descriptors}. Most of such studies use an extremely small dataset (<200 samples) coupled with selected elemental properties to achieve high accuracy ($R^2$>0.95) as evaluated by the random hold-out or cross-validation methods.  However, due to the high similarity (or redundancy) among the samples, these two evaluation methods tend to over-estimate the prediction performance \cite{xiong2020evaluating}. Another reason for the reported high accuracy is because the space group and the structure topologies of the samples in their dataset are all identical and the variation among the lattice structures is only due to variations of some element properties.

On the other hand, few studies have been reported to predict lattice constants of generic crystal materials with varying mole ratios or different number of elements. In 2017, Takahashi et al. \cite{takahashi2017descriptors} calculated the lattice constants of 1541 binary body centered cubic crystals using density functional theory and trained a prediction model using Support vector regression (SVR) and feature engineering based on atomic number, atomic radius, electronegativity, electron affinity, atomic orbital, atomic density, and number of valence electrons. They reported a $R^2$ accuracy of 83.6\%. In 2020, Liang et al. \cite{liang2020cryspnet} proposed a neural network called Cryspnet based on extended Magpie elemental descriptors \cite{magpieward2016general} for lattice constants prediction for materials with formulas of generic forms. They built a neural network model for each Bravais lattice type. For Cubic (P,I,F) materials, their models achieve $R^2$ scores of 0.85, 0.80, and 0.83. For other crystal systems, the $R^2$ regression performance range from 0.11 to 0.77, with increasing scores for Bravais lattice types with higher symmetry. These scores are much lower than the other reported performance in the perovskite studies \cite{nait2020prediction,zhang2020machine}, indicating the much greater challenge in solving the generic lattice constant prediction problem.

While choice of machine learning algorithms affect the prediction performance, it is found that the descriptors play a major role in lattice parameter prediction. Fundamentally, any factor that contributes to the change of lattice constants can be added to the descriptor set of the prediction model. A large number of descriptors have been used for lattice constant prediction even though some of them may be specific to the form of the chemical compositions. For perovskite lattice constant prediction, the following descriptors have been used: valence electron\cite{verma2009lattice,chonghe2003prediction}, ionic radii(which reflects bond lengths)\cite{verma2009lattice,nait2020prediction}, tolerance factor($t=\frac{r_A+r_X}{\sqrt2.(r_B+r_X)} $) calculated from ionic radii of the A-site and B-site cations and rX, the ionic radius of the anion)\cite{jiang2006prediction}, electronegativity\cite{brik2014lattice,brik2011modeling,majid2009lattice}, oxidation state\cite{majid2009lattice}, ionic charge\cite{kumar2009lattice}. In term of ionic radii, there are several ways to combine the values of the  component elements such as the sum, differences, or ratios. In the structure based lattice constant prediction\cite{williams2020deep}, both graph representations and 2D fingerprints have been used by measuring (di, de) at each point on the surface and then binning them into a 2D histogram. In lattice constant prediction of binary body centered cubic crystals, Keisuke Takahashi et al. \cite{takahashi2017descriptors} recommended seven descriptors for predicting lattice constant a including atomic number of element A and B, density, atom orbital of elements A and B, difference in electronegativity between A and B, atomic orbital B+difference in electronegativity between A and B. When trained with 1541 samples, their Support vector regression model achieved mean $R^2$ score of 83.6\% via cross-validation and maximum error of 4\% when compared to experimentally determined lattice constants. However it is not obvious how to extend their two-element oriented descriptors to materials with three or more elements.



In this work, we focus on the lattice edge length ($a,b,c$) prediction problem for generic crystal materials. Compared to previous work, our datasets are much larger (10-100 times), enabling us to achieve high performance for generic lattice parameter prediction. Our model have achieve exceptionally high accuracy for cubic systems with $R^2$ reaching 0.979, of which the materials have a single edge length $a$ as their lattice parameter. Cubic crystals consist of 18,325
 or 14.63\% of all the 125,278 crystals deposited in the Materials project database as of September 2020. Using a dataset with 18,325 samples of cubic crystals in Materials Project\cite{jain2013commentary}, we develop a random forest model using a set of novel descriptors for generic crystal materials lattice edge length prediction of which the number of elements and the mole ratios are not stereotyped. Our experiments show that our MLatticeABC algorithm achieves the much better performance compared with previous methods with an $R^2$ reaching as high as to 98\% for cubic materials.

Our contributions can be summarized as follows:

\begin{itemize}
    \item We propose a new descriptor set for generic lattice constant $a$,$b$,$c$ prediction of crystal materials.
    \item We conduct extensive experiments with different combinations of descriptors and transfer learning strategies and evaluate and compare the performances of different machine learning algorithms.
    \item Our experiments show that our MLatticeABC algorithm based on Random forest achieves state-of-the-art prediction performance in generic lattice edge length prediction.
\end{itemize}

\section{Materials and Methods}

\subsection{Descriptors}
In this work, we focus on developing lattice constant prediction models from materials composition only with the goal for downstream crystal structure prediction\cite{hu2020contact}. Such composition based models are also needed for large-scale screening of hypothetical materials composition datasets generated by generative machine learning models\cite{dan2020generative}. 

In the baseline model for generic lattice constant prediction\cite{liang2020cryspnet}, Magpie descriptors plus a few new descriptors have been used. The Magpie predictor set \cite{magpieward2016general} is based on calculating the mean, average deviation, range, mode, minimum, and maximum of a set of elemental properties (weighted by the fraction of each element in the composition). The element properties included in the Magpie descriptor calculation include Atomic Number,Mendeleev Number, Atomic Weight, Melting Temperature, Periodic Table Row \& Column, covalent radius, electronegativity, the number of Valence e- in each Orbital (s, p, d, f, total), the number of unfilled e-in each orbital (s, p, d, f, total), ground state band gap energy, ground state magnetic moment. Additionally, they have added the following descriptors including Stoichiometry p-norm (p=0,2,3,5,7), Elemental Fraction, Fraction of Electrons in each Orbital, Band Center, Ion Property (possible to form ionic compound, ionic charge).

Compared to the previous studies of lattice constant predictions for perovskites \cite{javed2007lattice}, one major difference of the generic lattice parameter prediction problem is the varying number of elements and the different mole ratios in the compositions of materials. For example, the following formulas are all included in our dataset which have different number of elements and stoichiometries: Sn\textsubscript{4}S\textsubscript{4}, Pr\textsubscript{20}S\textsubscript{32}, Ge\textsubscript{4}Sb\textsubscript{4},Sm\textsubscript{64}Cd\textsubscript{16}Ni\textsubscript{16},Mg\textsubscript{4}Co\textsubscript{16}O\textsubscript{32}. While descriptors based on ionic radii, electronegativity, ionic charges and so on have been defined for lattice constant prediction of perovskites, they need to be extended or adjusted to be applicable for formulas with varying number of elements and different mole ratios.

In this work, we start with the Magpie descriptors and the baseline descriptors in CryspNet\cite{liang2020cryspnet}, and propose MLatticeABC, a random forest lattice parameter ($a/b/c$) prediction model with a new set of descriptors. The feature set of our model include the enhanced Magpie features as described in \cite{liang2020cryspnet} (which includes the Magpie descriptors plus stoichiometry, valence orbitals, ion properties, band center, number of elements, and element fractions) and an additional descriptor set that characterizes the distribution of atom numbers of all elements in the formula. This element\_atom\_no\_stat descriptor set includes the statistics of the number of atoms of each element in the formula, from which we calculate the minimum, maximum, mean, and variance. It also includes the number of total atoms in the unit cell, which is a key feature that we found to be important for lattice edge length prediction. We also compare the performance of our model with the Roost model\cite{goodall2019predicting} in which a whole graph based graph neural network is used to extract descriptors for composition based property prediction. This model has achieved outstanding results in composition based formation energy prediction \cite{bartel2020critical}.

    




\subsection{Machine learning models: Random forest(RF), deep neural networks(DNN) and Gaussian Process Regressors(GPR)}

In this study, we combine different descriptors with deep neural networks, Random Forest, and Gaussian process regressor to identify the best model for lattice constant prediction.

We use the Random Forest \cite{liaw2002classification} to create  lattice  prediction models. RF is a supervised ensemble learning algorithm that constructs a multitude of many decision trees at training time and outputs the average of the regression values of the individual trees. RF algorithms have demonstrated strong prediction performance when combined with composition features in our previous studies \cite{cao2019convolutional}. In our RF regression models, we set the number of trees to be 50,criterion to 'mse'. This algorithm was implemented using the Scikit-Learn library in Python 3.6.

\begin{figure}[ht]
  \centering
  \includegraphics[width=0.6\linewidth]{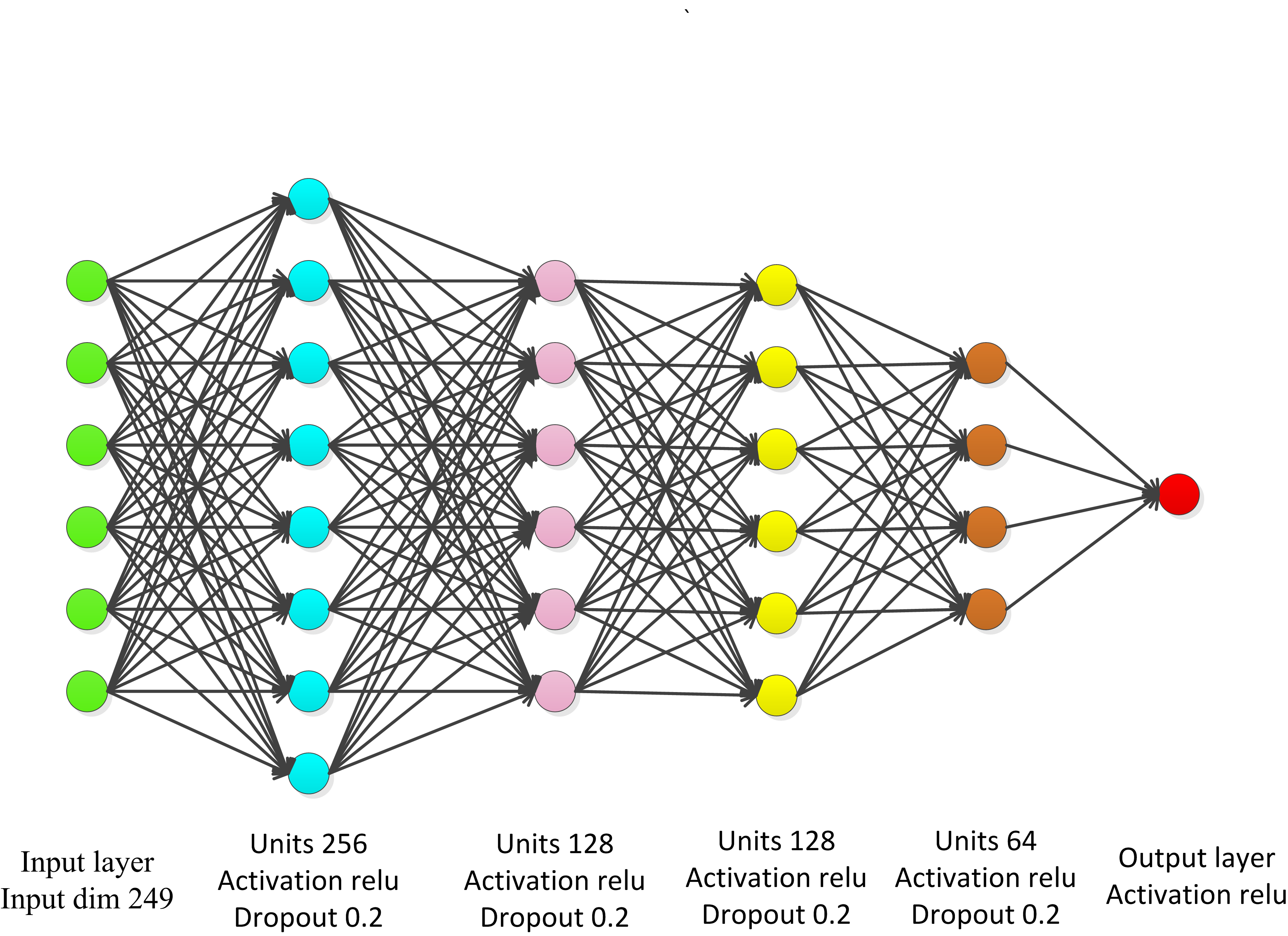}
  \caption{Architecture of the deep neural network.}
  \label{fig:NN}
\end{figure}

The deep neural network as shown in Figure\ref{fig:NN} is composed of 5 fully connected layers with 249, 256, 128, 128, 64, and 1 nodes from the input layer to the output layer. The action functions for those layers are relu. After each layer except the last layer, there is a dropout layer with 0.2 as the drop rates. The learning rate is set as 0.002. The batch size is 100. Standard Gradient Descent (SGD) is used as the optimizer. The network parameters are optimized manually to achieve the best performance with reasonable amount of trial-and-error fine-tuning.

As comparison, we also include Gaussian process regressor (GPR) models in our evaluation, which is a nonparametric, Bayesian algorithm for regression with several unique benefits: working well on small datasets and having the ability to provide uncertainty estimation on the predictions. We use the Scikit-learn implementation of the GPR with the WhiteKernel and RBF kernel. The noise level of the WhiteKernel is 1, the length scale of the RBF is 1.


\subsection{Evaluation criteria}

We evaluate model performance by splitting a whole dataset into 90\% as training and 10\% as testing and then train the model and make predictions. This process is repeated ten times to get the average performance scores. As a standard regression problem, the following three performance criteria have been used to compare the performance of different models, including  the root mean square error (RMSE), mean absolute error (MAE), determination of coefficient ($R^2$). 

\begin{equation}
R M S E=\sqrt{\frac{1}{n} \sum_{i=1}^{n}\left(a_{i}^{exp}-a_{i}^{pred}\right)^{2}}
\end{equation}

\begin{equation}
M A E=\frac{1}{n} \sum_{i=1}^{n}\left|a_{i}^{exp}-a_{i}^{pred}\right|
\end{equation}

\begin{equation}
R^{2}=\left(\frac{\sum(a_{i}^{exp} a_{i}^{pred})-\left(\sum a_{i}^{exp}\Sigma a_{i}^{pred}\right)/n}{\sqrt{\left[\sum a_{i}^{{exp}^{2}}-\frac{\left(\sum a_{i}^{{exp}^{2}}\right)}{n}\right]\left[\sum a_{i}^{{pred}^{2}}-\frac{\left(\Sigma a_{i}^{{pred}^{2}}\right)}{n}\right]}}\right)^{2}
\end{equation}

where $a_{i}^{exp}$  denotes experimental lattice constant $a/b/c$ of sample i; $a_{i}^{pred}$ is the predicted lattice constant value of $a/b/c$ for sample i.

\section{Experiments}

\subsection{Datasets}

\begin{figure}[hb!]
	\centering
	\begin{subfigure}{.33\textwidth}
		\includegraphics[width=\textwidth]{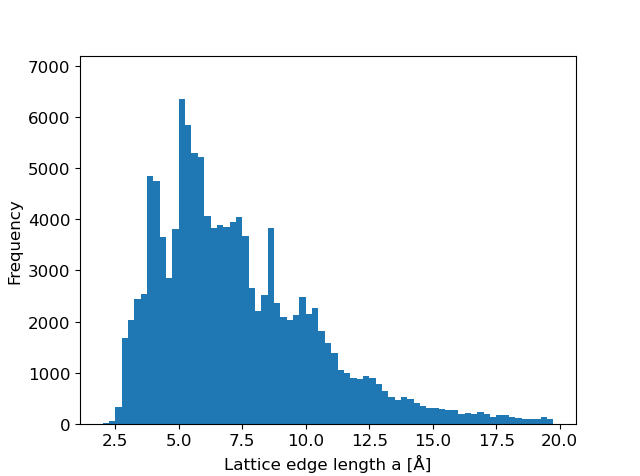}
		\caption{Histogram of $a$}
		\vspace{3pt}
	\end{subfigure}
	\begin{subfigure}{.33\textwidth}
		\includegraphics[width=\textwidth]{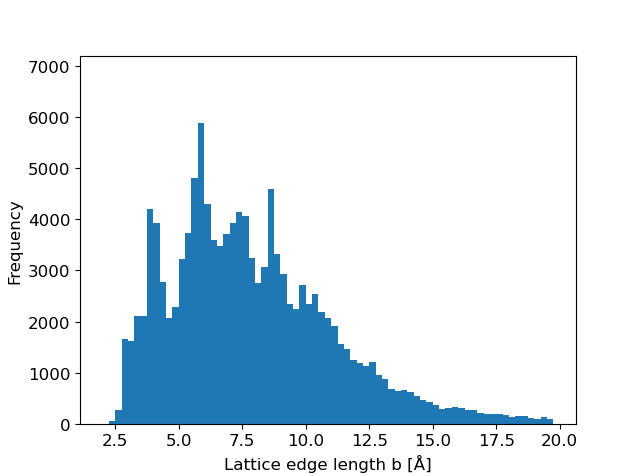}
		\caption{Histogram of $b$}
		\vspace{3pt}
	\end{subfigure}
	\begin{subfigure}{.33\textwidth}
		\includegraphics[width=\textwidth]{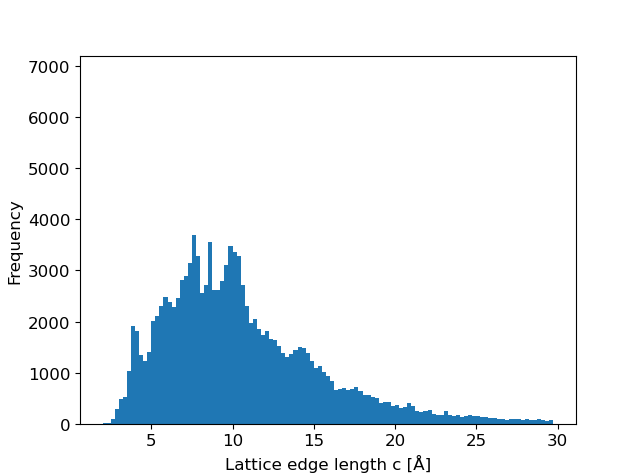}
		\caption{Histogram of $c$}
	\end{subfigure}

	\caption{Histogram of the lattice constant(edge length $a/b/c$) distribution.}
	\label{fig:histogram_abc}
\end{figure}

We use the Materials Project API to download the crystal lattice information of known inorganic materials deposited in the Materials Project database \url{http://www.materialsproject.org}. We observe that direct use of MPDataRetrievav.get\_dataframe function with structure.lattice.a/b/c/alpha/beta/gamma leads to inconsistent dataset because these lattices are the primitive lattice parameters. However, for cubic crystals, about 13,000 of the crystals' primitives are smaller units with $\alpha$/$\beta$/$\gamma$ equal to 60 degree instead of 90 degree while the remaining 5000 or so cubic crystals have a cubic unit cell representation with 90 degrees for $\alpha$/$\beta$/$\gamma$. To address this inconsistency issue, we instead download all the structural cif files of the crystals in the conventional\_standard format, and read the lattice parameters from them. In total, we get 125278 materials on September 10,2020 from the Materials Project database, which we call it as MP dataset.
These materials can be divided into seven groups by their crystal system types with corresponding sample numbers: Triclinic (15297), Monocinic (29872), Orthorhombic(26801), Tetragonal(14654), Trigonal(11086),  Hexagonal(9243), and Cubic(18325). For these materials, the distribution of the lattice cell lengths $a,b,c$ are shown in Figure\ref{fig:histogram_abc}. It can be found that most of the values of $a$ and $b$ are distributed between 2.5 \AA{} and 17.5 \AA{} and peaks are around 6 \AA{}. The values of lattice parameter c are mainly distributed between 2 and 25 \AA{} with more flat distribution.

\begin{figure}[ht]
  \centering
  \includegraphics[width=0.6\linewidth]{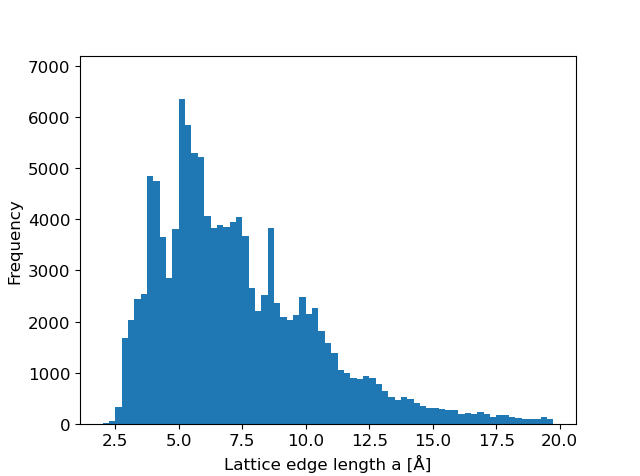}
  \caption{Histogram of the lattice constant(edge length $a$) distribution for cubic materials}
  \label{fig:histogram_a}
\end{figure}

As one of the major focus of this study, we also show the lattice parameter $a$ of cubic materials in Figure \ref{fig:histogram_a}. The distribution is similar to the overall distribution of $a$ in Figure \ref{fig:histogram_abc} with two peaks around 4 and 6 \AA. In addition, we show the distribution of samples in terms of element number in the crystals in our overall dataset in Table \ref{elementno_dist}. It is found that most of the samples are ternary and quaternary, and binary materials.


\begin{table}[h]
\begin{center}
\caption{The distribution of samples in terms of element number in the crystal.}
\label{elementno_dist}
\begin{tabular}{|l|c|c|c|c|c|c|c|c|c|}
\hline
The number of elements in each crystal & 1      & 2       & 3       & 4       & 5      & 6      & \textgreater{}=7 \\ \hline
The number of crystal                  & 713    & 19123   & 58592   & 34866   & 10200  & 1601   & 183              \\ \hline
Percentage                             & 0.57\% & 15.26\% & 46.77\% & 27.83\% & 8.14\% & 1.28\% & 0.15\%           \\ \hline
\end{tabular}
\end{center}
\end{table}











\subsection{Results}

\subsubsection{Prediction performance of MLatticeABC for cubic materials}

\begin{figure}[h]
	\centering
	\begin{subfigure}{.45\textwidth}
		\includegraphics[width=\textwidth]{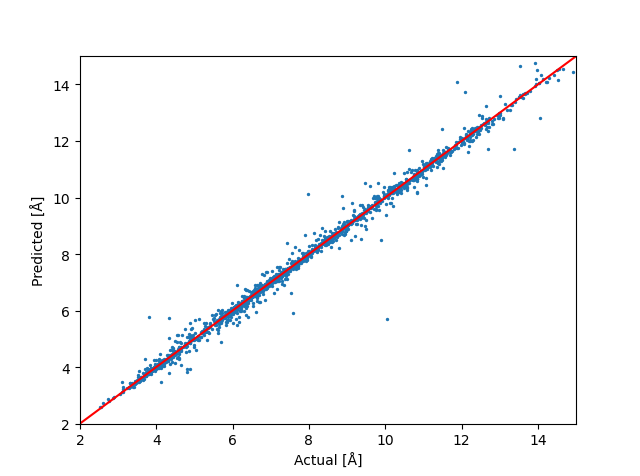}
		\caption{Parity plot of predicted lattice parameter $a$ of cubic materials}
		\vspace{3pt}
	\end{subfigure}
	\begin{subfigure}{.45\textwidth}
		\includegraphics[width=\textwidth]{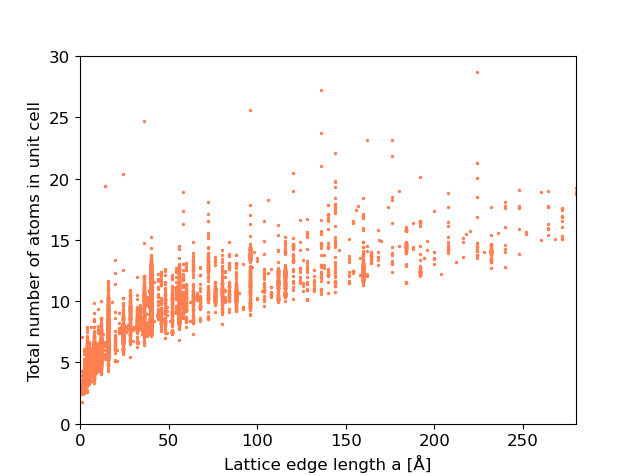}
		\caption{Correlation }
		\vspace{3pt}
	\end{subfigure}
		\caption{High prediction performance for lattice parameter $a$ prediction for cubic materials and the key parameter, the total number of atoms. (a) True and predicted lattice constant parameter $a$ by MLatticeABC where the descriptors include the enhanced Magpie feature + total atom number. (b) Correlation of total atom number with lattice $a$ of cubic materials.}
	\label{performance1}
\end{figure}


We run 10 times of random hold-out experiments and report the average and standard deviation of performance measures of these runs in terms of $R^2$, RMSE, and MAE. For our random forest model with the enhanced mapgie+total atom number feature, the average $R^2$, RMSE and MAE for unit cell length $a$ prediction for the cubic crystals is 0.979,0.138 and 0.416, which are much better than the baseline results reported in \cite{liang2020cryspnet}. In their report(Table 4), the length $a$ prediction performances are reported for three cubic Bravais lattices with $R^2$ of 0.80, 0.83, and 0.85 respectively. Figure \ref{performance1}(a) shows the parity plot of one of our hold-out experiments for lattice $a$ prediction of cubic materials, which shows it fits well with the predictions.

 To understand why the number of atoms in a material is so critical to the prediction of lattice edge length $a$ for cubic systems, we plot the parity plot of $a$ against the total atom number in Figure\ref{performance1}. It shows that there is a strong correlation of the number of atoms in the unit cell and the lattice edge length $a$ for cubic crystals. Due to the varying volume of atoms of different elements, for crystals with a specific lattice parameter $a$, the number of atoms also varies.

\subsubsection{Prediction performance of MLatticeABC for all crystal materials}

We conduct extensive experiments to evaluate our model performance in predicting the $a,b,c$ over all the crystal systems in the downloaded MP dataset using the same 10-repeats random cross-validation evaluation approach. The results are shown in Table\ref{abc_result}. It is found our RF model can achieve up to $R^2$ of 0.979 in lattice parameter a prediction for cubic materials using only composition as input. The performances over other crystal systems except Monoclinic are also good ranging from 0.77 to 0.892. In general, the higher the symmetry, the better the prediction performance. It is also interesting to find that the prediction performance over lattice $b$ and $c$ are all lower than those on lattice $a$ with all $R^2$ scores are between 0.489 and 0.77.

\begin{table}[h]
\begin{center}
\caption{ Prediction performance of MLatticeABC  in terms of $R^2$ score for $a,b,c$ over different crystal systems}
\label{abc_result}
\begin{tabular}{|l|c|c|c|c|c|}
\hline
Crystal system & train set size & test set size & $a$     & $b$     & $c$     \\ \hline
Cubic         & 16492          & 1833          & 0.979$\pm$0.016 &              &       \\ \hline
Hexagonal     & 8318           & 925           & 0.892$\pm$0.019 &              & 0.760$\pm$0.061 \\ \hline
Trigonal      & 9977           & 1109          & 0.843$\pm$0.025 &              & 0.705$\pm$0.065 \\ \hline
Tetragonal    & 13188          & 1466          & 0.846$\pm$0.024 &              & 0.685$\pm$0.044 \\ \hline
Orthorhombic  & 24120          & 2681          & 0.770$\pm$0.019 & 0.579$\pm$0.074 & 0.638$\pm$0.028 \\ \hline
Monoclinic    & 26884          & 2988          & 0.520$\pm$0.021 & 0.507$\pm$0.016 & 0.489$\pm$0.029 \\ \hline
Triclinic     & 13767          & 1530          & 0.800$\pm$0.022 & 0.771$\pm$0.014 & 0.650$\pm$0.056 \\ \hline
\end{tabular}
\end{center}
\end{table}

\begin{figure}[htb!]   \centering   \includegraphics[width=0.6\linewidth]{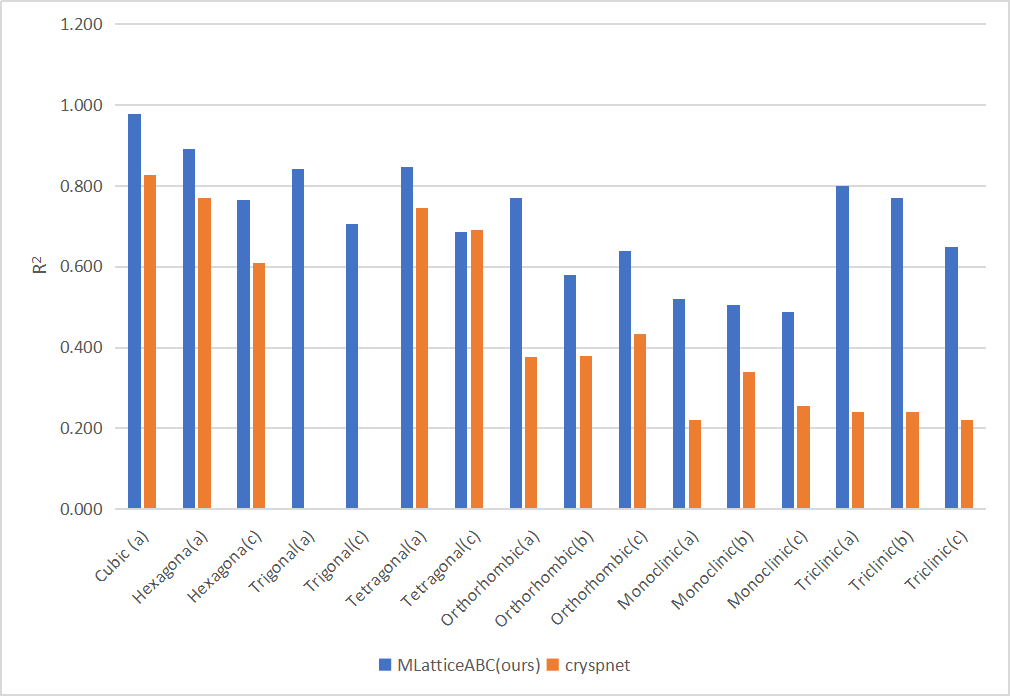}   \caption{Performance comparison of MLatticeABC and cryspnet for lattice constant ($a,b,c$) prediction in terms of $R^2$. For cryspnet, the shown scores are the average for crystal systems with multiple Bravais lattice types. Since there is no simple way to map trigonal crystals to Cryspnet lattice type models, their performances on this crystal system is not reported.}   
\label{fig:compare_crysp} 
\end{figure}

We also compare our model performance over all lattice systems with those of the baseline algorithm, cryspnet\cite{liang2020cryspnet}, which is a neural network based model for generic lattice parameter, crystal system, and space group prediction using the composition as the only input. In this approach, Matminer library were used to generate descriptors, which are then fed to the neural network to predict the Bravais lattice, space group and lattice parameters. The performance comparison is shown in Figure\ref{fig:compare_crysp}.  We find that for all the crystal systems, the performance of MLatticeABC is better, with significant improvements for Triclinic crystals. The performance gap may be due to the fact that cryspnet does not use global composition feature such as the total atom number and the statistics of atoms of different elements.

\FloatBarrier
\subsubsection{Prediction performance of MLatticeABC with data augmentation}

It is also known that crystals of different crystal/lattice systems have some special patterns in terms of their lattice parameters. Table\ref{latticepattern} shows the relationship of lattice parameters of the crystals. In previous work on generic lattice parameter prediction \cite{liang2020cryspnet}, a separate model is trained for each Bravais lattice type. However, we wonder if the training samples from other lattice systems can help to improve the performance. For example, can we build more accurate model for lattice parameter $a$ prediction of cubic materials by including other rectangular cuboid crystals as training samples since they are strongly related? To verify this hypothesis, we did the following experiments by first splitting the cubic samples into 90\% training and 10\% testing sets. And then we use the 90\% training plus the samples of tetragonal and orthorhombic to train the model and evaluate their performance on the original 10\% test samples. We repeat this 10 times to get the average scores. The descriptor set and the machine learning model are the same as in MLatticeABC. The results are shown in Table \ref{dataaugumentation}. The crystal system in the parenthesis is the crystal type of the test set. We find that the performance of the ML model with extended training samples always decreases for lattice constant $a$ prediction for both cubic (from line 2 to line 7) and orthorhombic crystals (from line 8 to line 11). This means the data augmentation strategy which train models with samples from other lattice systems have an adverse effect for the prediction performance and it is wise to train separate lattice parameter prediction models for different crystal systems using samples of that specific crystal systems only.

\begin{table}[h]
\begin{center}
\caption{ lattice parameter relationships for materials of different lattice systems. }
\label{latticepattern}
\begin{tabular}{|l|l|l|l|}
\hline
Crystal system & Edge lengths                                          & Axial angles                                    & Space groups \\ \hline
Cubic          & $a=b=c$                                                 & $\alpha=\beta=\gamma=90$                             & 195-230      \\ \hline
Hexagonal      & $a=b$                                                   & $\alpha=\beta=90, \gamma=120$                        & 168-194      \\ \hline
Trigonal       & $a=b\neq c$                         & $\alpha=\beta=90, \gamma=120$                        & 143-167      \\ \hline
Tetragonal     & $a=b\neq c$                         & $\alpha=\beta=\gamma=90$                             & 75-142       \\ \hline
Orthorhombic   & $a\neq b\neq c$ & $\alpha=\beta=\gamma=90$                             & 16-74        \\ \hline
Monoclinic     & $a\neq c$ & $\alpha=\gamma=90, \beta\neq 90$ & 3-15         \\ \hline
Triclinic      & all other cases                                       & all other cases                                 & 1-2          \\ \hline
\end{tabular}
\end{center}
\end{table}

\begin{table}[h]
\begin{center}
\caption{ Prediction performance comparison for lattice parameter $a$ of cubic materials with data augmentation}
\label{dataaugumentation}
\begin{tabular}{|l|c|c|c|}
\hline
\multicolumn{1}{|c|}{Descriptors}                        & Train sample no & Test sample no & $a$     \\ \hline
(Cubic)                                           & 16492           & 1833           & 0.979$\pm$0.016                 \\ \hline
(Cubic)+Tetragonal                                & 31146           & 1833           & 0.971$\pm$0.010                 \\ \hline
(Cubic)+Orthorhombic                              & 43293           & 1833           & 0.916$\pm$0.023                 \\ \hline
(Cubic)+Tetragonal+Orthorhombic                   & 57947           & 1833           & 0.934$\pm$0.009                 \\ \hline
(Cubic)+Hexagonal+Trigonal                        & 36821           & 1833           & 0.960$\pm$0.019                 \\ \hline
(Cubic)+Tetragonal+Orthorhombic+Hexagonal+Trigona & 78276           & 1833           & 0.933$\pm$0.009                 \\ \hline
(Orthorhombic)                                    & 24120           & 2681           & 0.770$\pm$0.019                 \\ \hline
(Orthorhombic)+Cubic                              & 42445           & 2681           & 0.731$\pm$0.032                 \\ \hline
(Orthorhombic)+Cubic+Tetragonal                   & 57099           & 2681           & 0.718$\pm$0.024                 \\ \hline
(Orthorhombic)+Cubic+Tetragonal+Trigona           & 68185           & 2681           & 0.701$\pm$0.027                 \\ \hline 
\end{tabular}
\end{center}
\end{table}

\FloatBarrier

\subsubsection{Performance comparison of different algorithms} 

\begin{figure}[h]
	\centering
	\begin{subfigure}{.45\textwidth}
		\includegraphics[width=\textwidth]{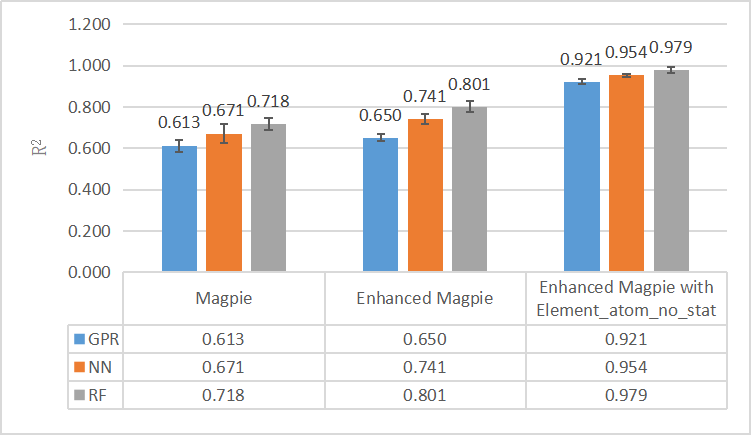}
		\caption{Performance comparison in terms of $R^2$}
		\vspace{3pt}
	\end{subfigure}
	\begin{subfigure}{.45\textwidth}
		\includegraphics[width=\textwidth]{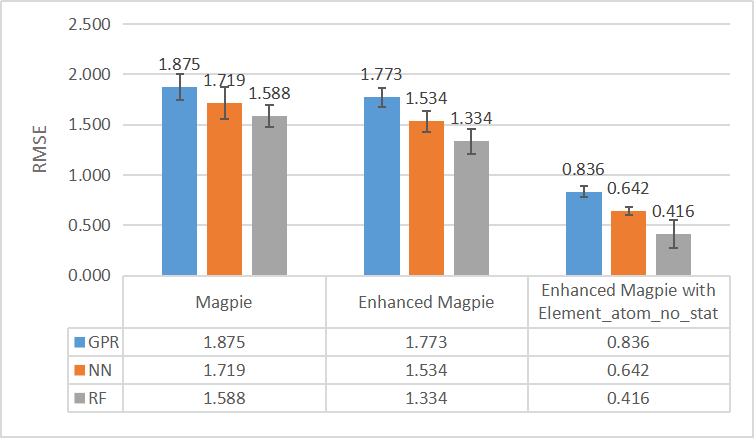}
		\caption{Performance comparison in terms of RMSE}
		\vspace{3pt}
	\end{subfigure}
	\caption{Performance comparison of different algorithms for lattice constant $a$ prediction}
	\label{fig:performanceParityplot}
\end{figure}

\begin{figure}[h]
	\centering
	\begin{subfigure}{.33\textwidth}
		\includegraphics[width=\textwidth]{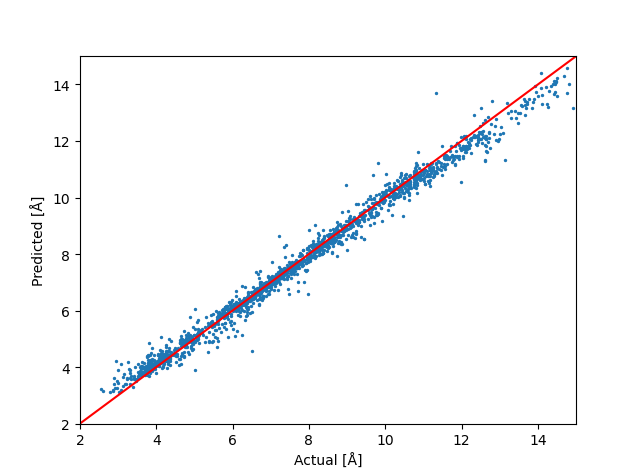}
		\caption{ DNN+ complete descriptor set}
		\vspace{3pt}
	\end{subfigure}
	\begin{subfigure}{.33\textwidth}
		\includegraphics[width=\textwidth]{figures/rf_true_pred.png}
		\caption{RF+ complete descriptor set}
		\vspace{3pt}
	\end{subfigure}
	\begin{subfigure}{.33\textwidth}
		\includegraphics[width=\textwidth]{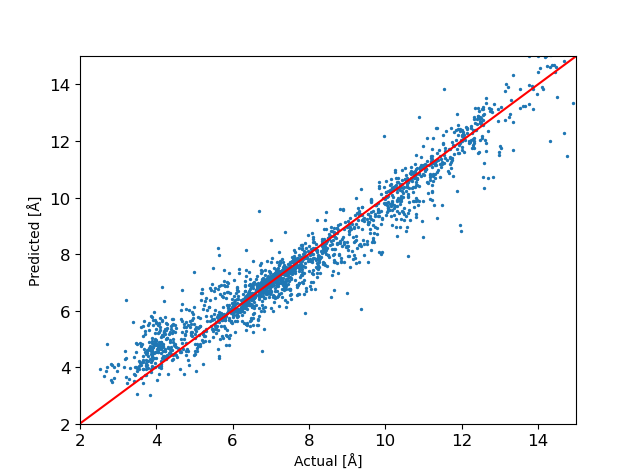}
		\caption{ GPR+ complete descriptor set}
		\vspace{3pt}
	
	\end{subfigure}	
\caption{Parity plots of different algorithms for lattice constant ($a$) prediction for cubic materials}
	\label{fig:performance-compare}
\end{figure}

Many different machine learning algorithms and descriptors have been used for lattice parameter prediction. Here we evaluate how DNN, Random Forest, and Gaussian process regressor perform with different feature sets. We use the cubic crystal dataset and repeat 10 times of hold-out experiments. The performance comparison is shown in Figure 
\ref{fig:performanceParityplot} and Table \ref{table:performance-compare}, which shows the baseline algorithm performance as reported in \cite{liang2020cryspnet}. First from Figure\ref{fig:performanceParityplot}, we find that the performances of RF model in terms of both $R^2$ and RMSE are always better than those of NN, which are better than the performance of GPR. Second, it is found that the enhanced Magpie features can clearly boost the prediction performance. Moreover, the figure shows that the performance improvements due to the inclusion of atom number is significant. The parity plots in Figure\ref{fig:performance-compare} further shows that the RF with our descriptor set has the best performance.

\begin{table}[!htb] 
\begin{center}
\caption{ Prediction performance comparison with baselines. }
\label{table:performance-compare}
\begin{tabular}{|l|c|c|c|}
\hline
\textbf{Algorithm} & \textbf{$R^2$} & \textbf{MAE} & \textbf{RMSE}
\\ \hline
\hline
GPR with Magpie                        & 0.613$\pm$0.029 & 1.140$\pm$0.048 & 1.870$\pm$0.132 \\ \hline
DNN with Magpie                        & 0.671$\pm$0.044 & 0.948$\pm$0.054 & 1.719$\pm$0.157 \\ \hline
RF with Magpie                         & 0.718$\pm$0.028 & 0.800$\pm$0.033 & 1.588$\pm$0.108 \\ \hline
GPR with Enhanced Magpie               & 0.650$\pm$0.017 & 1.084$\pm$0.033 & 1.773$\pm$0.093 \\ \hline
DNN with Enhanced Magpie               & 0.741$\pm$0.024 & 0.769$\pm$0.029 & 1.534$\pm$0.104 \\ \hline
RF with Enhanced magpie                & 0.801$\pm$0.027 & 0.517$\pm$0.028 & 1.334$\pm$0.122 \\ \hline
GPR with Element\_atom\_no\_stat + Enhanced Eagpie & 0.921$\pm$0.012 & 0.500$\pm$0.009 & 0.836$\pm$0.058 \\ \hline
DNN with Element\_atom\_no\_stat + Enhanced Eagpie & 0.954$\pm$0.007 & 0.336$\pm$0.038 & 0.642$\pm$0.041 \\ \hline
RF with Element\_atom\_no\_stat + Enhanced Eagpie  & 0.979$\pm$0.016 & 0.138$\pm$0.010 & 0.416$\pm$0.142 \\ \hline
\end{tabular}
\end{center}
\end{table}

\section{Discussion}

Our extensive experiments show that the prediction performance is strongly affected by the available dataset including its size and sample distribution and also the descriptors used. Our data augmentation results show that including the samples from other lattice systems can deteriorate the prediction performance for the crystal system specific ML models, which is contrary to the common practice of data augmentation or transfer learning. Here it is more appropriate to train individual models using only samples of that lattice system.

To gain further understanding of how descriptor combinations affect the model performance, we compare performance of different feature combinations using a set of ablation experiments. We evaluate the RF models only on the cubic system for simplicity. From Table \ref{table:accuracy_atomsiteno}, we first find that the  element\_atom\_no\_stat descriptor set is better than both Magpie and enhanced Magpie as proposed in \cite{liang2020cryspnet} with a $R^2$ score of 0.890 compared to 0.801 and 0.718 of Magpie and Enhanced Magpie respectively. When combined with Enhanced Magpie, the prediction performance of RF is greatly improved by including the element atom number statistics in the unit cell.

\begin{table}[!htb] 
\begin{center}
\caption{ RF Prediction performance comparison with different descriptor combinations. }
\label{table:accuracy_atomsiteno}
\begin{tabular}{|l|c|c|c|}
\hline
\textbf{Feature sets} & \textbf{$R^2$} & \textbf{MAE} & \textbf{RMSE} \\ \hline
\hline
Magpie               & 0.718$\pm$0.028 & 0.800$\pm$0.033 & 1.588$\pm$0.108 \\ \hline
Enhanced Magpie      & 0.801$\pm$0.027 & 0.517$\pm$0.028 & 1.334$\pm$0.122 \\ \hline
Element\_atom\_no\_stat          & 0.890$\pm$0.018 & 0.670$\pm$0.020 & 0.972$\pm$0.069 \\ \hline
Element\_atom\_no\_stat + Magpie & 0.972$\pm$0.015 & 0.142$\pm$0.009 & 0.500$\pm$0.135 \\ \hline
Element\_atom\_no\_stat + Enhanced Magpie & 0.979$\pm$0.016                 & 0.138$\pm$0.010                 & 0.416$\pm$0.142                 \\ \hline
\end{tabular}
\end{center}
\end{table}

We also compare our model performance with Roost\cite{goodall2019predicting}, a powerful graph neural network based prediction model using only composition as the input. In this approach, a composition formula is represented as a complete graph, which are then fed to a message passing graph neural network for feature extraction and regression. This approach has been shown to have strong extrapolation performance compared to other composition features \cite{bartel2020critical} in formation energy prediction. However, we find that since roost does not use global composition feature such as the total atom number, its performance is not good compared to our MLatticeABC. It achieves a $R^2$ score of only 0.78 for lattice length $a$ prediction of cubic materials.


\section{Conclusion}

Computational prediction of crystal unit cell lattice constants has wide applications in property investigation and crystal structure prediction. While previous studies have focused on models for specialized chemical systems with the same structure (a single space group) and identical anonymous formula and trained with small datasets, this study aims on developing a generic lattice constant predictor for crystal materials with more than 125,278 samples and 18,325 samples for the cubic system. We propose MLatticeABC, a random forest machine learning model with a new feature set combined with the standard composition features such as Magpie descriptors for effective lattice constant (edge length) prediction. Extensive standard and ablation experiments show that our random forest models with the new descriptors can achieve a high prediction performance with $R^2$ as high as 0.98 for the cubic system covering more than 18,000 samples. These machine learning models along with our easy-to-use open source code at \url{https://github.com/usccolumbia/MLatticeABC} 
can contribute to downstream tasks such as materials property prediction, materials screening, and crystal structure prediction. 

\section{Contribution}
Conceptualization, J.H. and Y.L.; methodology, Y.L. and J.H.; software, Y.L. and J.H.; validation, Y.L. and J.H.;  investigation, Y.L.,J.H., R.D. and W.Y.; resources, J.H.; writing--original draft preparation, J.H. and Y.L.; writing--review and editing, J.H and Y.L.; visualization, Y.L.,J.H, R.D; supervision, J.H.;  funding acquisition, J.H.

\section{Acknowledgement}
Research reported in this work was supported in part by NSF under grant and 1940099 and 1905775 and by NSF SC EPSCoR Program under award number (NSF Award OIA-1655740 and GEAR-CRP 19-GC02). The views, perspective, and content do not necessarily represent the official views of the SC EPSCoR Program nor those of the NSF.

\section{Data Availability}
The data that support the findings of this study are openly available in Materials Project database \cite{jain2013commentary} at \url{http://www.materialsproject.org}.

\bibliography{references}

\begin{thebibliography}{10}

\bibitem{mccaffrey1973electronic}
Joseph~W McCaffrey, James~R Anderson, and Dimitrios~A Papaconstantopoulos.
\newblock Electronic structure of calcium as a function of the lattice
  constant.
\newblock {\em Physical Review B}, 7(2):674, 1973.

\bibitem{pozzo2011thermal}
Monica Pozzo, Dario Alfe, Paolo Lacovig, Philip Hofmann, Silvano Lizzit, and
  Alessandro Baraldi.
\newblock Thermal expansion of supported and freestanding graphene: lattice
  constant versus interatomic distance.
\newblock {\em Physical review letters}, 106(13):135501, 2011.

\bibitem{wang2017lattice}
Xiaotian Wang, Zhenxiang Cheng, Rabah Khenata, Yang Wu, Liying Wang, and
  Guodong Liu.
\newblock Lattice constant changes leading to significant changes of the
  spin-gapless features and physical nature in a inverse heusler compound
  zr2mnga.
\newblock {\em Journal of Magnetism and Magnetic Materials}, 444:313--318,
  2017.

\bibitem{jiang2006prediction}
LQ~Jiang, JK~Guo, HB~Liu, M~Zhu, X~Zhou, P~Wu, and CH~Li.
\newblock Prediction of lattice constant in cubic perovskites.
\newblock {\em Journal of Physics and Chemistry of Solids}, 67(7):1531--1536,
  2006.

\bibitem{liang2020cryspnet}
Haotong Liang, Valentin Stanev, A~Gilad Kusne, and Ichiro Takeuchi.
\newblock Cryspnet: Crystal structure predictions via neural network.
\newblock {\em arXiv preprint arXiv:2003.14328}, 2020.

\bibitem{hu2020contact}
Jianjun Hu, Wenhui Yang, Rongzhi Dong, Yuxin Li, Xiang Li, and Shaobo Li.
\newblock Contact map based crystal structure prediction using global
  optimization.
\newblock {\em arXiv preprint arXiv:2008.07016}, 2020.

\bibitem{hu2020distance}
Jianjun Hu, Wenhui Yang, and Edirisuriya M~Dilanga Siriwardane.
\newblock Distance matrix based crystal structure prediction using evolutionary
  algorithms.
\newblock {\em arXiv preprint arXiv:2009.13955}, 2020.

\bibitem{nait2020prediction}
Menad Nait~Amar, Mohammed~Abdelfetah Ghriga, Mohamed El~Amine Ben~Seghier, and
  Hocine Ouaer.
\newblock Prediction of lattice constant of a2xy6 cubic crystals using gene
  expression programming.
\newblock {\em The Journal of Physical Chemistry B}, 124(28):6037--6045, 2020.

\bibitem{williams2020deep}
Logan Williams, Arpan Mukherjee, and Krishna Rajan.
\newblock Deep learning based prediction of perovskite lattice parameters from
  hirshfeld surface fingerprints.
\newblock {\em The Journal of Physical Chemistry Letters}, 2020.

\bibitem{chonghe2003prediction}
Li~Chonghe, Thing Yihao, Zeng Yingzhi, Wang Chunmei, and Wu~Ping.
\newblock Prediction of lattice constant in perovskites of gdfeo3 structure.
\newblock {\em Journal of Physics and Chemistry of Solids}, 64(11):2147--2156,
  2003.

\bibitem{javed2007lattice}
Syed~Gibran Javed, Asifullah Khan, Abdul Majid, Anwar~M Mirza, and J~Bashir.
\newblock Lattice constant prediction of orthorhombic abo3 perovskites using
  support vector machines.
\newblock {\em Computational materials science}, 39(3):627--634, 2007.

\bibitem{takahashi2017descriptors}
Keisuke Takahashi, Lauren Takahashi, Jakub~D Baran, and Yuzuru Tanaka.
\newblock Descriptors for predicting the lattice constant of body centered
  cubic crystal.
\newblock {\em The Journal of chemical physics}, 146(20):204104, 2017.

\bibitem{cao2019convolutional}
Zhuo Cao, Yabo Dan, Zheng Xiong, Chengcheng Niu, Xiang Li, Songrong Qian, and
  Jianjun Hu.
\newblock Convolutional neural networks for crystal material property
  prediction using hybrid orbital-field matrix and magpie descriptors.
\newblock {\em Crystals}, 9(4):191, 2019.

\bibitem{majid2011predicting}
Abdul Majid, Asifullah Khan, and Tae-Sun Choi.
\newblock Predicting lattice constant of complex cubic perovskites using
  computational intelligence.
\newblock {\em Computational Materials Science}, 50(6):1879--1888, 2011.

\bibitem{zhang2020machine}
Yun Zhang and Xiaojie Xu.
\newblock Machine learning lattice constants for cubic perovskite a2xy6
  compounds.
\newblock {\em Journal of Solid State Chemistry}, page 121558, 2020.

\bibitem{moreira2007comment}
Roberto~L Moreira and Anderson Dias.
\newblock Comment on “prediction of lattice constant in cubic perovskites”.
\newblock {\em Journal of Physics and Chemistry of Solids}, 68(8):1617--1622,
  2007.

\bibitem{majid2010lattice}
Abdul Majid, Asifullah Khan, Gibran Javed, and Anwar~M Mirza.
\newblock Lattice constant prediction of cubic and monoclinic perovskites using
  neural networks and support vector regression.
\newblock {\em Computational materials science}, 50(2):363--372, 2010.

\bibitem{ubic2015lattice}
R~Ubic, K~Tolman, K~Talley, B~Joshi, J~Schmidt, E~Faulkner, J~Owens, M~Papac,
  A~Garland, C~Rumrill, et~al.
\newblock Lattice-constant prediction and effect of vacancies in aliovalently
  doped perovskites.
\newblock {\em Journal of alloys and compounds}, 644:982--995, 2015.

\bibitem{ubic2009prediction}
Rick Ubic and G~Subodh.
\newblock The prediction of lattice constants in orthorhombic perovskites.
\newblock {\em Journal of alloys and compounds}, 488(1):374--379, 2009.

\bibitem{majid2009lattice}
Abdul Majid, Muhammad~Farooq Ahmad, and Tae-Sun Choi.
\newblock Lattice constant prediction of a 2 bb’o 6 type double perovskites.
\newblock In {\em International Conference on Computational Science and Its
  Applications}, pages 82--92. Springer, 2009.

\bibitem{ahmad2019artificial}
Rashid Ahmad, Aqsa Gul, and Nasir Mehmood.
\newblock Artificial neural networks and vector regression models for
  prediction of lattice constants of half-heusler compounds.
\newblock {\em Materials Research Express}, 6(4):046517, 2019.

\bibitem{xiong2020evaluating}
Zheng Xiong, Yuxin Cui, Zhonghao Liu, Yong Zhao, Ming Hu, and Jianjun Hu.
\newblock Evaluating explorative prediction power of machine learning
  algorithms for materials discovery using k-fold forward cross-validation.
\newblock {\em Computational Materials Science}, 171:109203, 2020.

\bibitem{magpieward2016general}
Logan Ward, Ankit Agrawal, Alok Choudhary, and Christopher Wolverton.
\newblock A general-purpose machine learning framework for predicting
  properties of inorganic materials.
\newblock {\em npj Computational Materials}, 2(1):1--7, 2016.

\bibitem{verma2009lattice}
AS~Verma and VK~Jindal.
\newblock Lattice constant of cubic perovskites.
\newblock {\em Journal of alloys and compounds}, 485(1-2):514--518, 2009.

\bibitem{brik2014lattice}
Mikhail~G Brik, Andrzej Suchocki, and Agata Kaminska.
\newblock Lattice parameters and stability of the spinel compounds in relation
  to the ionic radii and electronegativities of constituting chemical elements.
\newblock {\em Inorganic chemistry}, 53(10):5088--5099, 2014.

\bibitem{brik2011modeling}
MG~Brik and IV~Kityk.
\newblock Modeling of lattice constant and their relations with ionic radii and
  electronegativity of constituting ions of a2xy6 cubic crystals (a= k, cs, rb,
  tl; x= tetravalent cation, y= f, cl, br, i).
\newblock {\em Journal of Physics and Chemistry of Solids}, 72(11):1256--1260,
  2011.

\bibitem{kumar2009lattice}
A~Kumar and AS~Verma.
\newblock Lattice constant of orthorhomic perovskite solids.
\newblock {\em Journal of alloys and compounds}, 480(2):650--657, 2009.

\bibitem{jain2013commentary}
Anubhav Jain, Shyue~Ping Ong, Geoffroy Hautier, Wei Chen, William~Davidson
  Richards, Stephen Dacek, Shreyas Cholia, Dan Gunter, David Skinner, Gerbrand
  Ceder, et~al.
\newblock Commentary: The materials project: A materials genome approach to
  accelerating materials innovation.
\newblock {\em Apl Materials}, 1(1):011002, 2013.

\bibitem{dan2020generative}
Yabo Dan, Yong Zhao, Xiang Li, Shaobo Li, Ming Hu, and Jianjun Hu.
\newblock Generative adversarial networks (gan) based efficient sampling of
  chemical composition space for inverse design of inorganic materials.
\newblock {\em npj Computational Materials}, 6(1):1--7, 2020.

\bibitem{goodall2019predicting}
Rhys~EA Goodall and Alpha~A Lee.
\newblock Predicting materials properties without crystal structure: Deep
  representation learning from stoichiometry.
\newblock {\em arXiv preprint arXiv:1910.00617}, 2019.

\bibitem{bartel2020critical}
Christopher~J Bartel, Amalie Trewartha, Qi~Wang, Alex Dunn, Anubhav Jain, and
  Gerbrand Ceder.
\newblock A critical examination of compound stability predictions from
  machine-learned formation energies.
\newblock {\em npj Comput Mater 6, 97}, 2020.

\bibitem{liaw2002classification}
Andy Liaw, Matthew Wiener, et~al.
\newblock Classification and regression by randomforest.
\newblock {\em R news}, 2(3):18--22, 2002.

\end{thebibliography}
\bibliographystyle{unsrt}

\end{document}